\documentclass[10pt,a4paper]{article}

\usepackage[utf8]{inputenc}
\usepackage[T1]{fontenc}
\usepackage{authblk} 
\usepackage{graphicx}
\usepackage{float}
\usepackage{listings}

\usepackage[hidelinks]{hyperref} 

\title{Identifying Hearing Difficulty Moments in Conversational Audio}

\author{\footnotesize Jack Collins, Adrian Buzea, Chris Collier, Alejandro Ballesta Rosen,\\ Julian Maclaren, Richard F. Lyon, Simon Carlile}
\affil{\footnotesize Google Research Australia \\
\texttt{\scriptsize \{jackcollins, adrianbuzea, czcollier, alejandrobr, jmaclaren, dicklyon, scarlile\}@google.com}
}

\date{} 

\begin{document}

\maketitle

\begin{abstract}
Individuals regularly experience \textit{Hearing Difficulty Moments} in everyday conversation. Identifying these moments of hearing difficulty has particular significance in the field of hearing assistive technology where timely interventions are key for realtime hearing assistance. In this paper, we propose and compare machine learning solutions for continuously detecting utterances that identify these specific moments in conversational audio. We show that audio language models, through their multimodal reasoning capabilities, excel at this task, significantly outperforming a simple ASR hotword heuristic and a more conventional fine-tuning approach with Wav2Vec, an audio-only input architecture that is state-of-the-art for automatic speech recognition (ASR).
\end{abstract}

\section{Introduction}
According to a recent study by the WHO, there are currently 1.5 billion people affected by hearing loss which could grow to 2.5 billion by 2050. For 430 million, this loss is severe and has a deleterious effect on quality of life, physical and mental health and annually costs around \$1 trillion (US) in treatment and lost productivity. \cite{who2021}

For this study, we define a \textit{Hearing Difficulty Moment} as an event when a participant in a conversation has difficulty understanding what was said. These moments are a regular part of day-to-day conversations, occurring more frequently for those with underlying hearing loss conditions. A truly intelligent hearing aid device will, among other features, be able to accurately detect when these moments occur in order to take proactive, mitigating measures for the wearer’s optimal listening experience.

Numerous studies have focused on predicting dialogue acts in conversations, primarily utilizing transcribed text as input. These approaches have employed various techniques, from Bayesian inference \cite{grau2004} and support vector machines \cite{tavafi2013} to convolutional and recurrent neural networks  \cite{lee2016}, \cite{khanpour2016}, \cite{liu2017}, achieving different degrees of success. Notably, no prior work that we could find has explored the use of audio input for dialogue act prediction.

Recent work has proposed novel methods for detecting hearing loss from facial expressions where moments of hearing difficulty are observed as expressions of discomfort or fatigue in conversants \cite{yin2024}. Motivated by this work in the visual domain, we explore how similar audio-only cues can be used for hearing difficulty detection.

This study is part of the AFHI (Australian Future Hearing Initiative), which focuses on new applications of AI and machine learning to develop listening and communications technologies that overcome the limitations of current hearing assistive technologies, in particular hearing aids, with a focus on more customised/precision hearing healthcare.

\section{Methods}
This research focuses on the detection of a subject's level of hearing difficulty at a specific time $t$, given the contextual audio waveform data from conversational segments preceding $t$.

We evaluate an ASR hotword heuristic solution, a Wav2Vec SFT model and audio language models (prompting and fine-tuning) as classifiers for this task.

\subsection{ASR Hotword Heuristic (Baseline)}
We employ a two-step process for establishing a simple baseline for this task. The first part uses Chirp 2, a specific implementation from Google's broader Universal Speech Model (USM) family of state-of-the-art speech models, \cite{google2024} to transcribe the audio. For the second part, we use a dictionary with hearing difficulty hotwords such as “what, pardon, sorry” to determine if the transcript contains any indication of hearing difficulty.

\subsection{Wav2Vec 2.0 transfer learning}
Wav2Vec 2.0 \cite{baevski2020} is a popular speech recognition model published by Meta. It is trained in a self-supervised manner, learning from unlabeled audio and is then fine-tuned on labeled transcriptions. The model achieves state-of-the-art results with limited labeled data and is relatively cheap from a computational point of view, especially compared to competitors such as Whisper.

We use the wav2vec2-base-960h model as a base model. It has approximately 95 million parameters and was pre-trained and fine-tuned on 960h of audio from the Librispeech dataset \cite{panayotov2015}. 

We replace the final layer of the Wav2Vec2.0 model with a standard 2-layer DNN classification head. The entire model is then trained using a learning rate of 1e-5 and a batch size of 8 with each batch containing an equal number of positive and negative examples as a method to combat the very imbalanced dataset. Freezing the model's trunk and training only the new head resulted in significantly lower performance compared to training all weights. In addition, no benefit was observed by first training the classification head with a frozen trunk and then unfreezing the trunk vs training the whole model with a very low learning rate from the start.

To enhance performance, a probabilistic data augmentation process is implemented during training. Each time a training example is sampled, Gaussian noise (amplitude uniformly sampled between 0.001 and 0.015), time stretching (with a fixed rate for all samples, uniformly sampled between 0.8 and 1.25), and pitch shifting (uniformly sampling between -4 and 4 semitones) are each applied with a 50\% likelihood. This, combined with random negative resampling per training batch, increased the model's robustness across diverse scenarios. The models typically converged within 30 to 50 training epochs.

\subsection{Prompted Audio Language Model}
To implement this classification task using the multimodal Gemini 1.5 Pro model, we repurpose the “P” and “N” tokens as our predicted classes. Specifically, the model is prompted with detailed instructions to analyze the audio signal for classification.

\begin{lstlisting}[breaklines]
You are an expert at analyzing if a speaker in a given conversation is having difficulties understanding or hearing at a given moment. Please consider the following factors:

  * **Non-semantic information:**  Assess the general tone and pitch expressed in the speakers voice. Is it predominantly strained, or at ease? The lombard effect describes the things to look out for when someone is struggling in conversation:
    - increase in phonetic fundamental frequencies
    - shift in energy from low frequency bands to middle or high bands
    - increase in sound intensity
    - increase in vowel duration
    - spectral tilting
    - shift in formant center frequencies for F1 (mainly) and F2
    - the duration of content words are prolonged to a greater degree in noise than function words
    - greater lung volumes are used
  * **Semantic information:**  Pay attention to what they are saying and any keywords which might indicate that they are struggling to understand something. Are they asking for clarifications? Common examples to look out for (not exhaustive):
    - What?
    - Can you repeat that?
    - I didn't catch that?
    - Huh?
    - Sorry?
  * **Subjectivity:**  Recognize that some experiences are inherently subjective. Focus on the speaker's experience rather than your personal opinions. Do you think they are having a moment of hearing difficulty?

  Use all of the context available but make your judgement only on if the current moment (ie. the end of the audio) is a hearing difficulty event or not.

  Answer only with "P" for POSITIVE meaning a hearing difficulty event or "N" for NEGATIVE meaning it isn't a hearing difficulty event. Do not include any other rationale or fluff in your response.
\end{lstlisting}

The prompt details both semantic and non-semantic information about a Hearing Difficulty Moment which the model is encouraged to pay attention to to inform its prediction. In the few-shot case, we present an equal number of (randomly drawn) positive and negative examples along with the corresponding ground truth label (Audio: [audio\_tokens], Label: P …). The audio segment for which we are generating the prediction is lastly appended to the prompt (Audio: [audio\_tokens], Label: ).

The next-token log probabilities of the “P” and “N” tokens are then retrieved to compute a relative confidence signal for the positive class. This method can yield a signal at each token step which predicts the probability of a Hearing Difficulty Moment at that point in the conversation. Such a signal can be observed over time as illustrated below.

\begin{figure}[h!]
    \centering
    \makebox[\textwidth][c]{\includegraphics[width=1.4\textwidth]{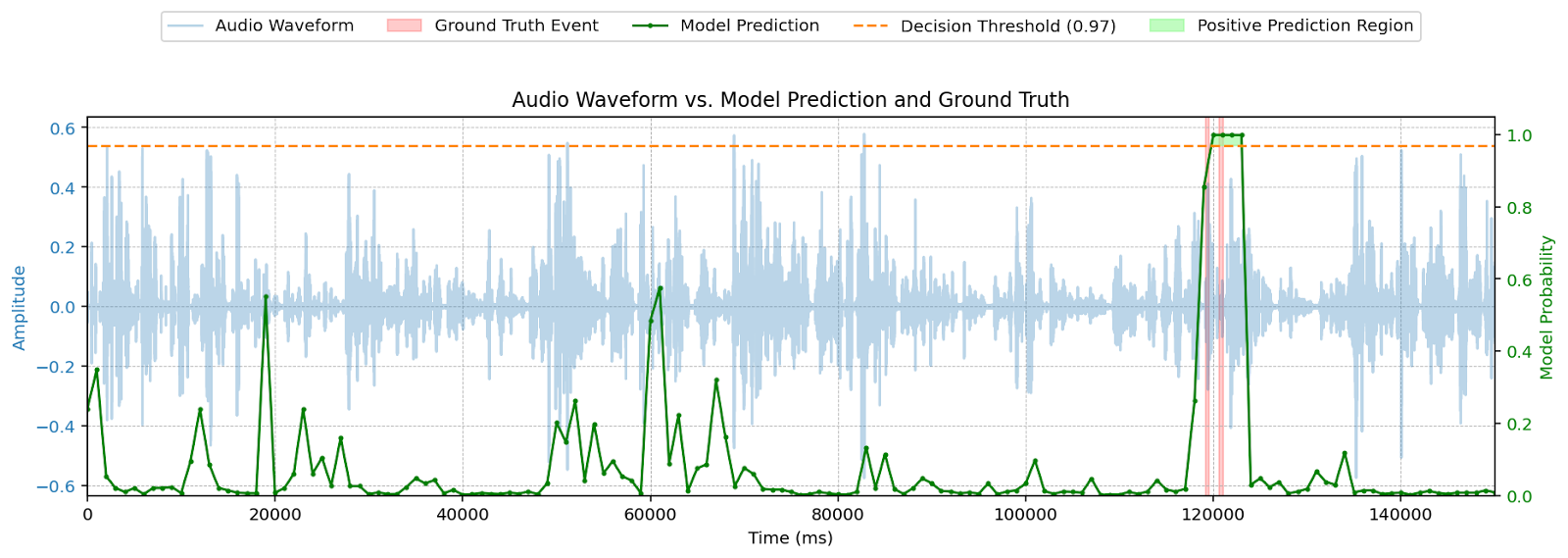}}
    \caption{Example of the continuous output from the Gemini 1.5 Pro 10-shot prompting method. The red shaded areas represent ground truth Hearing Difficulty Moments and the green line is the output probability of the “P” token obtained over multiple 4s windows of audio sampled every 1000ms.}
    \label{fig:fig1}
\end{figure}

In an alternative approach, the audio modality is withheld as an input, and a Chirp 2 text transcript is provided instead to Gemini 1.5 Pro \cite{gemini2024} in a 0-shot configuration along with a slightly modified prompt to remove the references to non-semantic audio cues, and to draw attention to the “transcript” rather than to the “audio”. This variation allows for observation of the uplift provided by the additional audio modality in this context.

\subsection{LoRA Fine-tuned Audio Language Model}
Low-Rank Adaptation (LoRA) \cite{hu2022} fine-tuning is a technique to efficiently adapt large pre-trained language models to specific tasks. Instead of fine-tuning all the model's parameters, LoRA freezes the original weights and introduces a pair of low-rank matrices to represent the changes. During training, only these smaller matrices are updated. This method significantly reduces the number of trainable parameters, leading to faster training and lower memory usage compared to full fine-tuning, while still achieving comparable or even better performance, particularly when using all available examples.

A multimodal Gemini 2.0 Flash model \cite{hassabis2024gemini} is fine-tuned over 14 epochs with a learning rate multiplier of 0.5, starting at an original learning rate of 1e-3 and a LoRA adapter size (rank) of 8. Training datasets contain an equal distribution of positive and negative examples; approximately 430 examples total. The models and training framework used are publicly available Google Cloud versions.

\section{Evaluation}

\subsection{Dataset}
Our dataset for identifying Hearing Difficulty Moments in conversational audio comprises 1,200 long conversations sourced from the Switchboard Dialog Act Corpus (SWDA) \cite{jurafsky1997} and Meeting Recorder Dialog Act Corpus (MRDA) \cite{shriberg2004} datasets. Both consist of 1,199 conversations, segmented into over 327,000 short utterances (typically a few seconds), containing different speakers (there is some overlap between speakers with MRDA) and different topics. 

Human annotation of dialogue acts in conversational speech is inherently subjective, leading to inconsistencies even among expert labelers. For example, in \cite{stolcke2000}, eight linguistics graduates achieved only 84\% agreement when labeling the SWDA dataset based solely on text.

Each utterance from MRDA and SWDA has already been assigned a human-annotated act tag (e.g. “Statement” or “Question”). The 2 datasets use slightly different schemes for annotation, which are mapped to the commonly used DAMSL(Dialog Act Markup in Several Layers) annotation scheme \cite{core1997}. 

The act tag of interest to us is “signal-non-understanding”, used to annotate utterances like “Huh?”, “What?” or “What was that?”, which indicates that the speaker did not understand what was just said. We find a total of 522 “signal-non-understanding” utterances across both datasets, which we manually refined to 298 by excluding ones which are not a fit for our use case (i.e. non-understanding due to difficulty hearing). The discarded utterances represent, for example, non-understanding due to semantics (participant understood the words said, but asks for clarification of the meaning or intent). 

To prevent data leakage, all utterances from a single conversation are grouped together, belonging entirely to either the training or validation set.

We have labels for conversational segments of varying lengths. To model this in the time domain and generate a continuous signal, we divide the conversation audio into fixed-length segments (frames) and propagate the original, irregular-length labels onto these segments.

\begin{figure}[ht]
    \centering
    \includegraphics[width=1\textwidth]{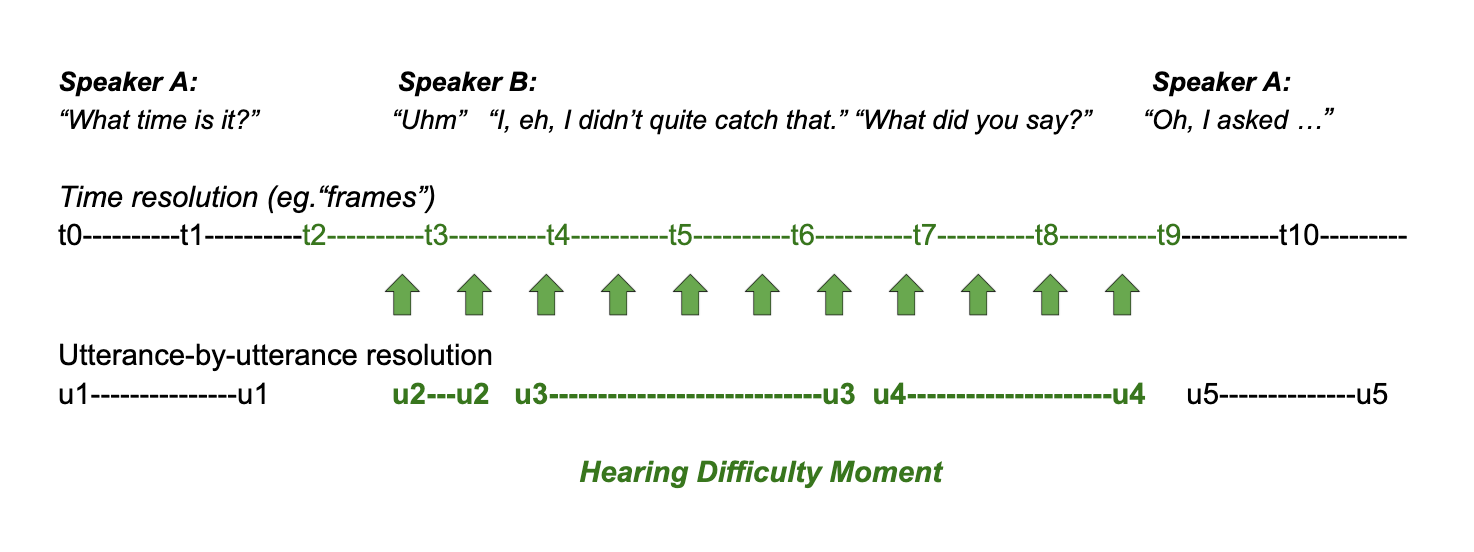}
    \caption{Hearing Difficulty Moments are propagated from the utterance level to the timestep level.}
    \label{fig:fig2}
\end{figure}

To sample positive Hearing Difficulty Moments in the time domain, we wish to construct examples that provide a reasonable amount of context before a moment takes place. To capture the relevant acoustic context, we opted for a four-second audio segment, a length consistent with established practices in sound classification research \cite{salamon2014}. The Hearing Difficulty Moments in our dataset are much shorter than that, with a mean length of 473ms.

\begin{figure}[ht]
    \centering
    \includegraphics[width=1\textwidth]{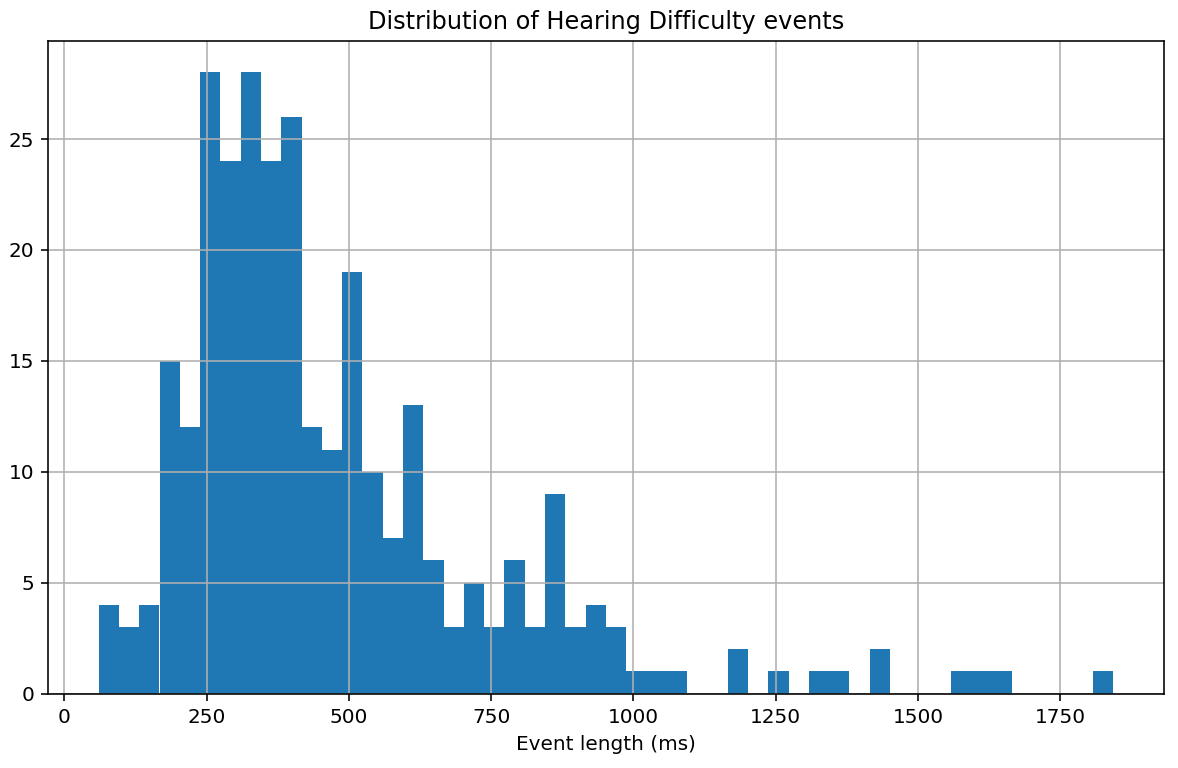}
    \caption{Distribution of lengths of Hearing Difficulty Moments in the dataset.}
    \label{fig:fig3}
\end{figure}

We randomly sample a positive timestep (where the event has been occurring already for a minimum of 0.4 seconds), taking the preceding 4 seconds of context audio. Similarly for negative events we randomly sample a negative timestep, taking the preceding 4 seconds of context audio, provided there is no overlap with a positive event.

\begin{figure}[ht]
    \centering
    \includegraphics[width=1\textwidth]{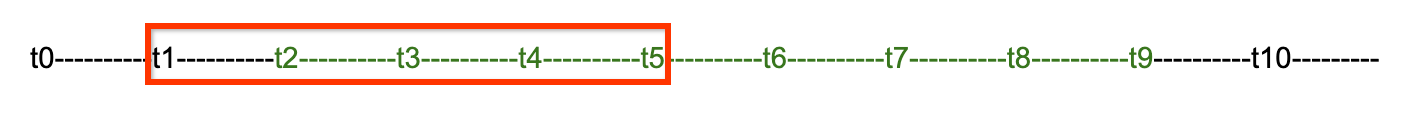}
    \caption{Examples consist of 4 seconds of sampled context audio under the described scheme.}
    \label{fig:fig4}
\end{figure}

\subsection{Monte Carlo Cross Validation}
We have a limited number of positive examples (298) and an abundance of negative examples.

We employ Monte Carlo cross-validation with 5 train/test splits. Each split randomly divides the conversations, allocating 80\% for training and 20\% for testing. As conversations may contain multiple positive events, the number of positive samples can vary between training and testing sets across different splits. For each positive instance, ten random negative instances are sampled from conversations within the same split, resulting in a 10:1 negative to positive ratio.

Monte Carlo cross validation helps mitigate variance arising from a single train/test split on our small dataset (e.g. simple examples in the test set). This approach is also beneficial for increasing the diversity of negative samples, as we randomly resample them for each split, unlike standard k-fold cross-validation which maintains a fixed set of both positive and negative examples.

\section{Results}

The ASR hotword heuristic served as the baseline comparison. The fine-tuned Wav2Vec classifier, representing a state-of-the-art ASR model, significantly outperformed the baseline, as anticipated. Remarkably, Gemini 1.5 Pro, in a zero-shot configuration without any historical examples of the task, achieved approximate performance parity with the Wav2Vec solution, based solely on the descriptive prompt provided. When prompted with just two randomly drawn examples (one positive, one negative), Gemini 1.5 Pro quickly becomes very competent at identifying these Hearing Difficulty Moments, showing a clear uplift over the Wav2Vec solution (F-score: 0.76) with an impressive F-score of 0.85. At 10-shot prompting (5 positive, 5 negative), Gemini 1.5 Pro shows a further performance uplift, reaching an F-score of 0.87. When Gemini 1.5 Pro was prompted without the audio tokens, relying solely on the Chirp 2 text transcript for information, a significant degradation in performance was observed. This approach had no observable uplift over the ASR hotword heuristic baseline.

\vspace{0.5cm}
\begin{tabular}{|l|c|r|}
\hline
Approach & Avg. F1 metric (5-fold MCCV) \\
\hline
ASR Hotword Heuristic (Baseline) & 0.39 \\
Gemini 1.5 Pro [text only] (0-shot) & 0.39 \\
Gemini 1.5 Pro [audio] (0-shot) & 0.75* \\
Wav2Vec 2.0 Transfer Learning & 0.76 \\
Gemini 2.0 Flash (LoRA Fine-Tuning) & 0.77 \\
Gemini 1.5 Pro (2-shot) & 0.85* \\
Gemini 1.5 Pro (10-shot) & 0.87 \\
\hline
\end{tabular}
\newline
\small * A statistically significant uplift was observed compared to the previous method ($p < 0.05$, one-tailed t-test with Nadeau and Bengio variance correction \cite{nadeau2003}).

\begin{figure}[H]
    \centering
    \includegraphics[width=1\textwidth]{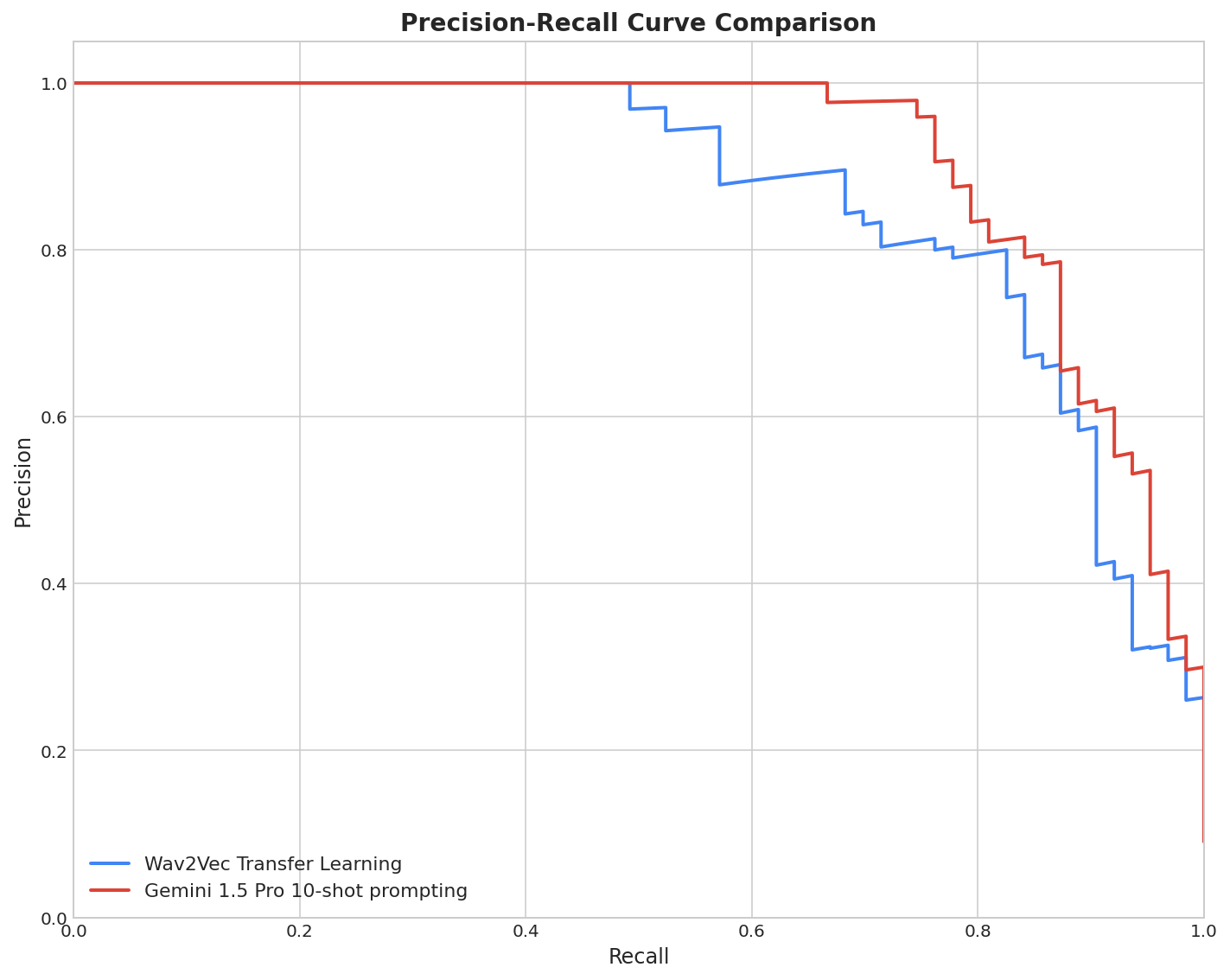}
    \caption{Precision-recall curves compared from a single fold for Wav2Vec and Gemini 1.5 Pro 10-shot prompting methods.}
    \label{fig:prcurve}
\end{figure}

\section{Discussion}
Leveraging audio language models for nuanced audio classification tasks involves a level of audio reasoning that goes well beyond ASR. We have established a Hearing Difficulty Moment detection task and shown that it is possible to achieve excellent results for this task with audio language models using both few-shot prompting and fine-tuning approaches, with the former showing significant uplift vs fine-tuning a prior state-of-the-art ASR model. We hypothesize that the uplift is explained by the cross-modality reasoning capabilities of the audio language models along with the large scale of training data used. The significant uplift seen in the audio modality methods vs the text-only method implies that non-semantic content is key in identifying these moments.

The training data in this study was composed of a 1:10 positive-to-negative label ratio. In practice, the occurrence of these positive events is likely to be orders-of-magnitude less frequent, depending on the setting and use case (according to \cite[p.~341]{stolcke2000} "signal-non-understanding" act tags have a frequency of 1 in 1,000 utterances in the Switchboard dataset \cite{jurafsky1997}). This would likely lead to a significant increase in false positives when switching to a more imbalanced serving distribution.

Given the nature of the training examples used in this study, we expect the methods described here to generalize well in controlled one-on-one and certain multi-speaker conversational settings. However, given the relative lack of variance in environment, lack of prolonged silences and other out-of-conversation dynamics which are largely missing in our training examples, the models described in this study would likely run into more serious limitations in ambient or “always on” settings. The integration of VAD-based heuristics could provide mitigation in such settings.

In \cite{yin2024}, a variation modelling approach was employed to determine if an individual is exhibiting facial expressions associated with the facial expressions of someone who indeed has underlying hearing loss. This opened new avenues for hearing loss screening outside of a clinical setting leveraging video data. In a similar vein, our study shows that by leveraging audio data, we can effectively track Hearing Difficulty Moments as they happen in conversations, potentially opening avenues for passively screening for the regularity with which individuals experience these moments in their daily lives. On the other hand, such a signal might also be used to optimize the hearing aid processing to favor speech sounds at the expense of other types of sounds given the apparent listening needs of the user in the moment.

We have focused here on predicting hearing difficulty at time $t$ given the 4 seconds of context that preceded time $t$. Real-world applications of this research may benefit from a slightly more relaxed variant of this task, where information for several seconds after time $t$ is also available at prediction time. This would likely lead to more powerful predictive power when identifying these events, at the cost of some prediction lag. There is also an open question regarding how the performance on this task varies with more or less preceding context than the fixed 4 seconds used in this study. Future work could further explore these variants of the task and the associated trade-offs. 

The study measured the performance of various methods on this task without taking the real-world latency and compute requirements of the solution into consideration. In particular, deploying audio language model solutions to edge devices currently presents significant challenges due to the size of these models. Future work will focus on distilling the knowledge of these solutions into smaller models which are suitable for edge deployments.

\end{document}